\documentclass[onecolumn, pra, nofootinbib,showpacs]{revtex4}

\usepackage{textcomp}
\usepackage{mathtools, ulem, graphicx}
\usepackage[export]{adjustbox}
\usepackage{sidecap}
\usepackage{stmaryrd}
\usepackage{natbib}
\usepackage{verbatim}
\newcommand{\e}{\begin{equation*}\begin{aligned}}
\newcommand{\ee}{\end{aligned}\end{equation*}}

\newcommand{\en}{\begin{equation}\begin{aligned}}
\newcommand{\een}{\end{aligned} \end{equation}}

\newcommand{\pfa}[2]{\frac{\delta #1}{\delta #2}}

\newcommand{\p}{\partial}
\newcommand{\pf}[2]{\frac{\p #1}{\p #2}}
\newcommand{\f}[2]{\frac{#1}{#2}}

\newcommand{\ra}{\rangle}
\newcommand{\la}{\langle}

\newcommand{\da}{\dagger}
\newcommand{\ma}{\mathcal}

\newcommand{\tr}{\text{Tr }}

\newcommand{\Q}{\left}
\newcommand{\W}{\right}

\newcommand{\pma}{\begin{pmatrix}}
\newcommand{\epma}{\end{pmatrix}}

\newcommand{\na}{\nabla}
\newcommand{\de}{\delta}
\newcommand{\ep}{\epsilon}

\newcommand{\psiu}{\psi(\tau,\vec{x})}
\newcommand{\psiud}{\psi^*(\tau,\vec{x})}
\newcommand{\tpsiu}{\psi(\vec{x})}
\newcommand{\tpsiud}{\psi^*(\vec{x})}

\newcommand{\tm}{\int^\beta_0 d \tau \,}
\newcommand{\xm}{\int d^2 \vec{x} \,}

\newcommand{\psiudd}{\psi^{\dagger}(\tau,\vec{x})}

\newcommand{\apsiu}{\psi '^{*}(\tau,\vec{x}')}
\newcommand{\apsiud}{\psi '(\tau, \vec{x}')}

\newcommand{\qpsipd}{\psi '^{*}(\tau,\vec{x}\,')}
\newcommand{\qpsip}{\psi '(\tau, \vec{x}\,')}

\begin{document}

\title{Path-integral Fujikawa's Approach to Anomalous Virial Theorems
and Equations of State for Systems with $\mathbf{SO(2,1)}$ Symmetry}
\author{Carlos R. Ord\'{o}\~{n}ez}
\affiliation{Department of Physics, University of Houston, Houston, TX 77204-5005}

\date{\today}

\email{cordonez@central.uh.edu}
%\begin{abstract}
%We give a brief introduction to the use of \LaTeX\ in the context of REVTeX~4.1.
%\end{abstract}


\begin{abstract}
We derive anomalous equations of state for nonrelativistic 2D complex bosonic fields with contact interactions, using Fujikawa's path-integral approach to anomalies and scaling arguments. In the process, we derive an anomalous virial theorem for such systems. The methods used are easily generalizable for other 2D systems, including fermionic ones, and of different spatial dimensionality, all of which share a classical $SO(2,1)$ Schr\"{o}dinger symmetry. The discussion is of a more formal nature and is intended mainly to shed light on the structure of anomalies in 2D many-body systems. The anomaly corrections to the virial theorem and equation of state
-- pressure relationship -- may be identified as the Tan contact term. The practicality of these ideas rests upon being able to compute in detail the Fujikawa Jacobian that contains  the anomaly.  This and other conceptual issues, as well as some recent developments, are discussed at the end of the paper.
\end{abstract}

\maketitle

\section{Introduction}

A quantum anomaly is the loss of a symmetry in a physical system as a consequence of rigorously defining its quantum description. They were discovered in particle physics, where the mandatory regularization and renormalization of quantum field theory spoiled some existing classical symmetries \cite{1a, 1b, 1c}. Anomalies may also occur in nonrelativistic systems, but they are less well known. R. Jackiw, one of the original co-discoverers of anomalies in relativistic systems, very clearly pointed out the possible existence of anomalies in quantum mechanics, specifically, in systems with attractive inverse square potentials, or in 2D systems with contact interactions (delta function) \cite{3}. These systems exhibit a classical $SO(2,1)$ conformal symmetry, which is lost after regularization and renormalization of the original theory is required to formally define the quantum theory (the method of self-adjoint extensions of operators is an alternative way of dealing with these systems \cite{a}. In what follows we are going to work within the regularization and renormalization framework). After Jackiw's seminal paper a number of authors developed these ideas further, concentrating mainly on formal aspects\footnote{See \cite{4,5} and references therein.}, although a nice application of these concepts was developed in molecular physics \cite{b}.  Recently, anomalies have been investigated in ultracold, diluted trapped 2D systems. In particular, following the seminal work by Pitaevskii \&  Rosch (P\&R) \cite{6}, J. Hofmann  showed the possible existence of anomalies for 2D trapped Fermi gasses with Hamiltonian \cite{8}

\en
H_{\text{osc}}=\int d^2x \left[\psi^\dagger_\sigma \frac{-\nabla^2}{2m}\psi_\sigma(x)+\frac{g}{m}\psi^\dagger_{\uparrow}\psi^\dagger_{\downarrow}
\psi_{\downarrow}\psi_{\uparrow}(x)  + \frac{m\omega^2_0 x^2}{2}\psi^\dagger_{\sigma}\psi_{\sigma}(x)\right] \equiv
H+\int d^2x \frac{m\omega^2_0 x^2}{2}\psi^\dagger_{\sigma}\psi_{\sigma}(x).
\een

Despite the presence of the trapping term  $\int d^2x \frac{m\omega^2_0 x^2}{2}\psi^\dagger_{\sigma}\psi_{\sigma}(x)$  in Eq. (1), which explicitly breaks the scale invariance of $H$, P\&R showed that the system possesses a \textit{classical} hidden $SO(2,1)$ symmetry, described in terms of the Lie-algebra generators $H$, $D$ (generator of dilations) and $C$, with
\begin{flalign}
&D=\int d^2x x_i m j_i(x), \\ \nonumber
&j_i=-i(\psi^\dagger\partial_i\psi-\partial_i\psi^\dagger \psi)/2m,\\
&C=\int d^2x \frac{mx^2}{2}\psi^\dagger_{\sigma}\psi_{\sigma}(x).
\end{flalign}

Hofmann noted that due to the need for regularization and renormalization of the quantum field theory, the classical scaling symmetry of the system is broken, which is manifested by the existence of a ground-state energy  $E_b$ (scale) in the (attractive) 2-body sector, with the ensuing alteration of the original Lie algebra. For instance, the anomaly-free commutation relation

\begin{equation}
[D,H]=2iH
\end{equation}

is modified due to the anomaly and it now includes an ``extra term''

\begin{equation}
[D,H]=2iH+\frac{i}{2\pi m}I,
\end{equation}

where  $I$ is the so-called Tan contact operator \cite{40a, 40b, 40c}

\en
I=\int d^2x\, g^2 \psi^\dagger_{\uparrow}\psi^\dagger_{\downarrow}\psi_{\downarrow}\psi_{\uparrow}(x).
\een

These algebraic modifications have experimental consequences  that can be measured. Hofmann calculated the anomalous frequency shift at $\delta \omega /\omega_0$ at zero temperature for the breathing modes and found that, in some cases, e.g., in the strongly interacting regime $\ln k_F a_{2D}\approx 0$ ($a_{2D}$ = 2D scattering length), the change could be of the order of 10\% compared to its symmetric classically predicted value. Other important effects due to the anomaly discussed by him were the shift in the ground-state energy

\begin{equation}
E_0=\langle H+\omega^2_0 C \rangle = 2\omega^2_0 \langle C \rangle - \frac{1}{4 \pi m}\langle I \rangle,
\end{equation}

and in the pressure of the untrapped system (classically, due to the $SO(2,1)$ symmetry, $\ma E-P=0$)

\begin{equation}
P=\mathcal{E}+\frac{\ma I}{4m},
\end{equation}

where $\ma E$ and $\ma I$ are the energy and contact density of the gas. Other similar issues and effects for ultracold Bose gasses have been discussed by Olshanii et al. \cite{42}, and a number of papers have been addressing theoretical and experimental aspects of the modifications of the physics of ultracold gases due to the presence of an anomaly \cite{50,51,52,53,54,55}. While much work has been and is being done, more work is needed in order to fully understand these anomalous effects \cite{58}. In particular, zero-temperature (anomaly) effects vs finite-temperature effects have to be further studied. Hofmann speculates that the experimental negative anomaly findings of the authors of \cite{7} may be explained as a consequence of finite-temperature corrections. The interconnection between zero and finite-temperature effects is indeed non-trivial. For instance, Chafin and Schafer recently computed the grand potential $\Omega$ to one loop for a 2D Fermi gas at finite temperature (same $H_{\text{osc}}$ as Eq.(1), but without the trapping term) \cite{62}. To deal with the infinities encountered they renormalized using the same renormalization scheme as in the description of the 2-body ground state (anomaly), and proceeded to calculate the shift to the free second virial coefficient to obtain \footnote{In a theory with classical scale invariance such as this one, the phase shifts should naively all be constant, and hence the second term of Eqs. (9) and (10) would be absent; but the existence of scaling anomalies introduces the bound state energy scale, and things change dramatically.}

\en \label{26}
\delta b_2=e^{\beta E_b}-2\int \f{dk}{k} \f{e^{-2\beta \epsilon_k}}{\left[\ln \left(\f{k^2}{E_b} \right) \right]^2+\pi^2}.
\een

This formula exactly coincided with the well-known Beth-Uhlenbeck formula \cite{65}

\en \label{27}
\delta b_2=e^{\beta E_b}+\f{1}{\pi} \int dk \left(\f{d\delta^{(0)}}{dk} \right)e^{-2\beta \epsilon_k}.
\een

$\delta^{(0)}$ is the only non-zero phase shift for the contact potential for this problem. $E_b$ is the finite energy of the ground state (the finiteness of $E_b$ is what produces the running of the coupling constant $\lambda$ of Eq. (1) as a function of the cutoff, which in turns allows for full renormalization of $\Omega$; that's why one sees the presence of $E_b$ in Eqs. (9) and (10)). Those equations show that there is a non-trivial comingling between zero-temperature (anomaly,$E_b$) and finite-temperature effects. It is therefore important to do a comprehensive study of regularization, renormalization issues in 2D systems, in both zero and finite-temperature sectors, in order to be able to interpret experimental results and develop a deeper understanding of these systems. In this note we will propose a framework that will combine Fujikawa's path-integral approach to anomalies \cite{10a,10b} and operator methods, with an emphasis on scaling properties and virial theorems, in the description of finite and zero-temperature anomaly effects in 2D ultracold diluted gasses. In order to better understand the role of the anomaly we will consider the untrapped homogeneous system of a $2D$ Bose gas. We hope that this new framework will help us address the issues
raised above. While the trapped system is the one that is of most
interest experimentally and what motivated this work, the anomalous
features are produced by short-distance physics and they are already
present in homogenous systems as several authors have  shown \cite{8,Ananos:2002id}. 
When possible, comments
about trapped systems will be made, but a complete study of those
systems will be left for future work. Just as Hofmann did using operator methods, in this paper we
will also find modifications to the symmetric equations (virial theorem and
pressure equation) due to anomalies, using
Fujikawa's approach.  The extra term in the equations
found here may, therefore, be identified with Tan's contact term,
basically Eq. (6) above. This will be seen in Eqs. (26), (35), and (53). However, the results in this paper are or a more formal and general
nature, and more work needs to be done in order to fully make these
identifications. The technical difficulty here is the explicit evaluation
of functional traces of operators, which requires delicate regularization.
Some progress has been made and it's mentioned in the comments and
conclusions.\\

\section{Virial Theorems In Quantum Field Theory and Anomalous Equation of State}

While quantum treatments of the classical virial theorem for particles have been available for some time, the analog theorems in quantum field theory are not as well known (see, however \cite{D} and \cite{G}).  The exception, in the case of nonrelativistic quantum fields (relevant for us), was the work of T. Toyoda et al. starting in the late 90s \cite{59,60,61}. They made use of the interpretation of the ``virial'' operator (below) as the generator of scale transformations and derived a series of virial theorems for finite-temperature, nonrelativistic field theories. We will not repeat their work here; instead, we will present a different, path-integral derivation of their results, at least in the case of a complex scalar field with a quartic interaction.  In their work, Toyoda et al. introduce an external potential to confine the system to a volume $V$.  We will not do so, and will simply assume the system has a finite, but very large volume $V$ so that the usual thermodynamics assumptions can be made. For the sake of clarity we chose this specific bosonic system, but the methods developed will be of a general nature for interacting fields with classical $SO(2,1)$ symmetry (the fermionic case can be treated in a similar fashion).

\subsection{ Virial Theorem}

Consider the partition function for our system:
\en \label{1}
\begin{split}
Z&=\text{tr}\, \Q( e^{-\beta(H-\mu N)}\W) \\
&=\int [d\psi^*][d\psi]e^{-S^{\mu,\beta}_E},
\end{split}
\een

 where 

\begin{multline} \label{2}
S^{\mu,\beta}_E=\int_0^\beta d\tau \int d^2\vec{x} \, \left(  \psiud \pf{}{\tau}\psiu+\f{1}{2}\nabla\psiud \cdot\nabla \psiu \right.   \\
+g \left(\psiud \psiu \right )^2 -\mu \left. \psiud \psiu\f{}{} \right).
\end{multline}

The fields obey the usual periodicity condition in Euclidean time $\tau$.  Following Toyoda et al., we're going to demand that under a rescaling of the coordinates  $\vec{x} \rightarrow \lambda \vec{x}$, the number operator be invariant:

\en \label{3}
\begin{split} 
\tpsiud \tpsiu d^2\vec{x}=\psi '^{*}(\vec{x} \,') \psi '(\vec{x}\,') d^2\vec{x}\,' ,
\end{split}
\een

which determines the transformation for the field variable (since the Euclidean time will play no role in what follows -- although it will later -- we will omit the time variable for the time being):

\en \label{4}
\begin{split}
\tpsiu \rightarrow \psi' (\vec{x}\,')=\lambda^{-1} \tpsiu .
\end{split}
\een
Next, we have to make assumptions about the transformation properties of the path-integral measure under the scaling transformations before proceeding further:

\begin{align} \label{5}
&\text{i)} [d\psi^*][d\psi]=[d\psi '^*][d\psi '] &\text{ (No Anomaly) }.\\ \label{6}
&\text{ii)} [d\psi^*][d\psi]=J^{\lambda^{-1}}[d\psi '^*][d\psi '] &\text{ } J^{\lambda^{-1}} \neq 1 \text{ (Anomaly) }. 
\end{align}

where $J^{\lambda^{-1}}$ is the Jacobian for a change of variables $(\psi,\psi^*) \rightarrow (\psi',\psi'^*); \text{ } J^{1}=1$; see appendix.\\

Consider case i) first.  Make the transformation  $\tpsiu=\lambda \psi ' (\lambda \vec{x})$.  By relabeling the integration variables, it's easy to see that only two terms change in the exponential in the path integral:

\en \label{7}
\f{1}{2} \int d^2 \vec{x}\,\nabla \tpsiud \cdot \nabla \tpsiu= \f{\lambda^2}{2}\int d^2 \vec{x}\,' \, \nabla' \psi '^{*}(\vec{x}\,')\cdot \nabla'\psi'(\vec{x}\,'),
\een

\en \label{8}
g \int d^2 \vec{x}\,\left(\tpsiud \tpsiu \right)^2=g \lambda^2 \int d^2 \vec{x}\,'\left( \psi '^{*}(\vec{x}\,') \psi '(\vec{x}\,') \right)^2.
\een

Make an infinitesimal scale transformation  $\lambda=1+\eta$.  Then the partition function becomes (the $\tau$ dependence has now been restored due to the $\tau$ integration):

\en \label{9}
\begin{split}
\int [d\psi^*][d\psi]e^{-S_E[\psi^*, \psi]}&=\int [d\psi '^*][d\psi']e^{-\Q(S_E[\psi '^*, \psi']+\delta S\W)} \\
&=\int [d\psi '^*][d\psi']e^{-S_E[\psi '^*, \psi']}
\times \left\{1-\f{2 \eta}{2}  \tm \xm ' \nabla' \qpsipd \cdot \nabla' \qpsip \right. \\
&- \left. 2 \eta g \tm \xm ' \left( \qpsipd \qpsip \right)^2  \right \}.
\end{split}
\een

Therefore, in the large volume limit \footnote{At this point there is no need to make a distinction between primed and unprimed variables.}

\en \label{10}
\left \langle \tm \xm \f{1}{2} \nabla \psiudd \cdot \nabla \psiu \right \rangle=-\left \langle \tm \xm g (\psiudd  \psiu)^2 \right \rangle,
\een

where

\en \label{11}
\langle A \rangle=Z^{-1} \text{ tr} \left(e^{-\beta(H-\mu N)} A \right).
\een

Now, using 

\en \label{12}
\begin{split}
\psiu&=e^{\tau \hat{K}}\tpsiu e^{-\tau \hat{K}}, \\
\psiudd&=e^{\tau \hat{K}} \psi^\dagger(\vec{x}) e^{-\tau \hat{K}}, \\
\hat{K}&=H-\mu N,
\end{split}
\een

we get

\en \label{13}
\begin{split}
\left \langle \nabla \psiudd \cdot \nabla \psiu \right \rangle&=Z^{-1} \text{ tr} \left ( e^{-\beta \hat{K}}e^{\tau \hat{K}} \nabla \psi^\da(\vec{x}\,) \cdot \nabla \psi(\vec{x}\,) e^{-\tau \hat{K}}  \right)\\
&=Z^{-1} \text{ tr} \left ( e^{-\tau \hat{K}} e^{-\beta \hat{K}}e^{\tau \hat{K}} \nabla \psi^\da(\vec{x}\,) \cdot \nabla \psi(\vec{x}\,)   \right ) \\
&=\left \langle \nabla \psi^\dagger(\vec{x}) \cdot \nabla \tpsiu \right \rangle \text{ : } \tau \text{-independent. }
\end{split}
\een

A similar treatment shows that the RHS of Eq. (\ref{10}) is $\tau$-independent, which produces an overall $\beta$ factor on both sides, hence finally giving

\begin{align} \label{14}
\left \langle \f{1}{2}\xm \nabla \psi^\dagger(\vec{x}) \cdot \nabla \tpsiu \right \rangle &=-\left \langle \xm g (\psi^\dagger(\vec{x}) \tpsiu)^2\right \rangle , \\ 
\label{15}
\text{or }\text{ }\text{ }\text{ }\text{ }\text{ } \langle H_0 \rangle &\equiv \langle K \rangle=-\langle H_{\text{int}}\rangle \equiv -\la V\ra \text{: Virial Theorem.}
\end{align}

(Recall that in the first-quantized version, $2\langle K \rangle=n \langle V \rangle$, for $V=\alpha r^n$ or for $V(\lambda r)=\lambda^n V(r)$, such as $\delta^2(\vec{r})$ ($n=-2$).)  \\

In case ii), we have to add the contribution due to the Jacobian, Eq. (\ref{6}). Following similar steps to the previous section, one obtains a modified virial theorem (see appendix):

\begin{comment}
\en \label{16} \nonumber
\begin{split}
\langle K \rangle &=- \langle V \rangle+\f{\ma A}{\beta}: \text{Anomalous Virial Theorem}, \quad \text{ } \quad \text{ } \quad \text{ } \quad \text{ } \quad \text{ } \quad \text{ } \quad \text{ } \quad \text{ } \quad \text{ } \quad \text{ } \quad \text{ } \quad (26a)
\end{split}
\een
\end{comment}

\begin{align}\label{16} \nonumber
\langle K \rangle &=- \langle V \rangle+\f{\ma A}{\beta}: \text{Anomalous Virial Theorem}, 
\tag*{(26a)}
\end{align}

where (Tr is a functional trace) $\ma A=- \tr \left[(1+\vec{x}\cdot \vec{\nabla})\delta(\tau-\tilde{\tau})\delta^2(\vec{x}-\tilde{\vec{x}}) \right]$ is the anomalous contribution from the Jacobian for the infinitesimal transformation $\tpsiu \rightarrow \psi'(\vec{x})=\tpsiu+\delta \tpsiu$, $\delta \tpsiu=-\eta(1+\vec{x}\cdot \vec{\nabla})\psi'(\vec{x})$.\\

Although the emphasis in this paper is the study of homogeneous systems, the derivation of the anomalous virial theorem including a trapping term of the form $\int^\beta_0\int d^2\vec{x}\, \f{m\omega^2_0}{2}\,x^2\, \psi^\da(\tau,\vec{x}\,)\psi(\tau,\vec{x}\,)$ along these lines would readily give the following version:\footnote{Notice the thermal expectation values in Eq. (26b) are computed using the full Hamiltonian of Eq. (1).}

\begin{align*}
\la K \ra=-\la V\ra+\f{\ma A}{\beta}+ \Q\la \int \f{m\omega^2_0}{2} d^2\vec{x}\,\psi^\da(\vec{x}\,)\psi(\vec{x}\,) \W \ra.
\tag*{(26b)}
\end{align*}

Comparison with Eq. (7) would suggest that the anomaly term $\ma A$ should be identified with Tan's contact in the $2D$ case (see also Eq. (2) in ref. \cite{40c}).\\

Evidently, an explicit evaluation of $\ma A$ would require careful regularization (and possible renormalization) of the functional trace (\cite{lin}; see below).\\

%If we compare Eq. ( ) with  Eq. (8), we see that the \textit{Tan contact term comes entirely from the Fujikawa Jacobian factor in the path integral, as it should be.} \\

\setcounter{equation}{26}
\subsection{Anomalous Equation of State}

Consider the partition function for the scaled system ($\vec{x} \rightarrow \vec{x}'=\lambda \vec{x}$, $\tau \rightarrow \tau$, $\psi(\tau,\vec{x}) \rightarrow \psi'(\tau,\vec{x}')=\lambda^{-1}\psi(\tau,\vec{x})$)

\en \label{1}
\begin{split}
Z^\lambda=\int [d\psi^*]'[d\psi]'e^{-S^{\mu,\beta,\lambda}_E},
\end{split}
\een

where

\begin{equation} \label{2}
\begin{split}
S^{\mu,\beta, \lambda}_E=\int_0^\beta d\tau \int d^2\vec{x}' \, \left(  \apsiud \pf{}{\tau}\apsiu+\f{1}{2}\nabla' \apsiud \cdot\nabla' \apsiu \right.   \\
+g \left(\apsiud \apsiu \right )^2-\mu \left. \apsiud \apsiu \f{}{} \right) \\
=
\int_0^\beta d\tau \int d^2\vec{x} \, \Bigg(  \psiud \pf{}{\tau}\psiu+\lambda^{-2} \Big( \f{1}{2}\nabla\psiud \cdot\nabla \psiu    \\
+g \left(\psiud \psiu \right )^2  \Big) -\mu  \psiud \psiu  \Bigg).
\end{split}
\end{equation}

Using ((11) and (16)) we get

\begin{flalign}
&Z^\lambda=J^\lambda \text{ tr} \Q(e^{-\beta\Q(\lambda^{-2}H-\mu N\W)} \W).\\
&(H=H_0+H_{\text{int}}=\f{1}{2} \xm \nabla \psi^\dagger(\vec{x}) \cdot\nabla \tpsiu+g \xm (\psi^\dagger(\vec{x}) \tpsiu)^2 ) \nonumber
\end{flalign}

$Z^\lambda$ must correspond to a change of volume in the system (not in temperature or pressure)\footnote{See \cite{virpap} for a graphical proof of this statement.}. In the large volume limit \cite{brown}

\begin{flalign}
&Z^\lambda=e^{-\beta \Omega^\lambda}=e^{\beta P A^\lambda}. \\ \nonumber
&A=2D\text{``Volume'' ; } \text{ }A^\lambda=\text{scaled 2D ``Volume''}=\lambda^2A.
\end{flalign}

For infinitesimal scalings $\lambda=1+\eta$, we get ($\Delta A=2\eta A$) from Eq. (29)

\en
Z^{1+\eta}=e^{\beta P(A+\Delta A)}=J^{1+\eta}\text{tr}\Q(e^{-\beta\Q((1+\eta)^{-2}H-\mu N\W)}\W).
\een

As before $J^{1+\eta}$ gives an infinitesimal contribution (see appendix)

\en
J^{1+\eta}=1+2\eta \tr\Q(\hat{\theta}_s \delta^3(x-y)\W),
\een

and

\en
\hat{\theta}_s =-\Q(1+\vec{x}\cdot \vec{\nabla} \W).
\een

For a  $2D$ volume $A$, under $\vec{x} \rightarrow \lambda \vec{x}$, $A \rightarrow A+\Delta A$, $\Delta A=2\eta A$. Expanding Eq. (31) in $2\eta$ on both sides, and collecting terms we get Toyoda's result with the anomalous contribution:

\en
PA=\la H_0 \ra +\la H_{\text{int}}\ra + \f{\tr\Q( \hat{\theta_s}  \delta^3(x-y) \W)}{\beta}=\la H \ra + \f{\tr\Q( \hat{\theta_s} \delta^3(x-y) \W)}{\beta},
\een

which can be rewritten as \footnote{In thermal equilibrium and in the large $V$ ($A$) limit, one expects $\text{Tr}(\hat{\theta}_s  \delta^3(x-y))$ to be independent of $\tau$ and $\vec{x}$ in our case (untrapped), which will provide an additional $\beta A$ factor that will cancel the similar term in Eq. (35). This indeed happens explicitly in the Jacobian calculation of paper \cite{lin}. If one includes the trapping term $\int d^2x \frac{m\omega^2_0 x^2}{2}\psi^\dagger_{\sigma}\psi_{\sigma}(x)$, one works with the anomalous density of Eq. (35).}

\en
\ma E-P=-\f{\text{Tr}\Q(\hat{\theta_s}  \delta^3(x-y) \W)}{\beta A}, \text{ }\text{ }\text{ } \ma E=\f{\la H \ra}{A}.
\een

Equation (35) clearly displays the anomaly effect on the thermal (many-body) system: if the symmetry is preserved ($J=1$), $\ma E=P$, as is expected. But in general, if $\tr\Q( \hat{\theta_s}  \delta^3(x-y) \W) \neq 0$, the equation is modified\footnote{While we have used the terminology of quantum anomalies, strictly
speaking, since Eq. (13) is not necessarily a symmetry of the action,
the formal Eq. (35) should be considered more of a ``modified'' equation
vis-\'{a}-vis Toyota el al. Furthermore, since the treatment in both this
and their work is basically formal, after proper regularization and
renormalization (presumably somewhat different in each case), both might
still give the same physical results, despite the difference in the
formal equations \cite{private}. In the case of an  actual
$SO(2,1)$  quantum anomaly (section D), the presence of a non-trivial
Jacobian $J\neq 1 $  would signify a real physical effect (symmetry breaking).
The path-integral methodology of the current approach is essentially
identical in both cases though, and hence the early use of the language
of anomalies.}. \\

Again, comparison with Eq. (8) and Eq. (3) of ref. \cite{40c} suggests as before that the term $\f{\text{Tr}\Q(\hat{\theta}_S \delta^3(x-y)\W)}{\beta}$ is to be identified as Tan's constact term in the $2D$ case.\footnote{This method also gives the results for $D=3$.} 

\subsection{Algebraic considerations}

Toyoda et al. used a mixture of scaling arguments and algebraic manipulations to deduce their equations of state. We will now explore the algebraic content of our formulation, and compare with theirs, as well as with Hofmann's. 

Using their confining potential method, they arrived at the following result (their formula (2.5) of reference [61] and the similar one for the 3D case in a previous paper differ from ours by an overall factor. They use isotropy arguments to derive their equations of state, which agrees with ours. We don't need to make any extra assumptions):

\en \label{29}
PA=-\f{1}{2} \langle [G,H] \rangle=\la H\ra ,  \text{ }\text{ }\text{ }\text{ }\text{ }H=H_0+H_{\text{int}}.
\een

$G$ is defined in Eq. (39) below. We can obtain the same formula within our formulation (in the non-anomalous case):

\en \label{30}
\begin{split}
Z^{1+\eta}&=\text{tr } e^{-\beta \left( (H+\delta H)-\mu N \right)}=\text{tr } e^{-\beta \left( H-\mu N \right)} \left(1-\beta \delta H \right)
\text{ }\text{ }\text{ }\text{ }\text{ }(\delta H \propto H, [H,N]=0)\\
&=Z^1-\beta \text{ tr}\left(e^{-\beta \left( H-\mu N \right)} \delta H\right)\\
&=Z^1+Z^1(2 \eta \beta)AP.
\end{split}
\een

Therefore,

\en \label{31}
AP=-\f{1}{2\eta} \langle \delta H \rangle.
\een

But, 

\en \label{32}
\delta H&=H^{1+\eta}-H^1=-2\eta H=[G,H]\eta, \\
G&=\text{``virial'' operator (generator of scalings)}=\f{1}{2}\int d^2\vec{x}\, \vec{x} \cdot \psi^\da \Q( \vec{\nabla}-\overset{\shortleftarrow}{\nabla}  \W)\psi,
\een

and we have used $[G,H]=-2H$, which can be explicitly verified for our type of Hamiltonian, and using the standard commutation relations for $\tpsiu$, $\psi^\da(\vec{x})$.\\

Using Eq. (\ref{32}) and Eq. (\ref{31}) we then arrive at Eq. (\ref{29}). This is a self-consistent derivation, if one ignores the zero-temperature $SO(2,1)$ symmetry of the system and its possible anomaly, but as we saw with the example of the calculation of the second virial coefficient of the 2D fermion gas, only at one's peril! Indeed, if we do assume that our virial $G$ is part of the set of 3 generators of $SO(2,1)$ that generates dilations, one has to take into account the anomaly in deriving consequences of the algebra (as Hofmann also showed).  The full dilation operator $D$ is related to $G$ as follows:

\en \label{33}
D=-2tH-iG.
\een                   

In the non-anomalous case, the commutator with $H$ is known to be (as part of the $SO(2,1)$ symmetry, where both time and space are transformed)

\en \label{34}
[D,H]=2iH.
\een

Eq. (\ref{34}) is obviously equivalent to  $[G,H]=-2H$. So Eq. (\ref{29}) can be written as

\en \label{35}
PA=-\f{i}{2} \langle [D,H] \rangle.
\een

In the anomalous case, $[D,H]=2iH+\text{extra terms}$ , which in Hofmann's case of the 2D fermion system gave the corrected equation, Eq. (8), and we of course also anticipate this in the bosonic case described here. 
Embedding the scaling operator $G$ into the full $SO(2,1)$ via Eq. (\ref{33}) is an expedited way to find a formal corrected Eq. for the relationship between pressure and energy density, but it's not too insightful in computational terms, which we would like to have in order to take such equation into an actual equation of state, virial expansion, etc. Therefore, a path-integral treatment similar to the one described above, but now performing both time and space scalings, $t \rightarrow \lambda^2t$, $\vec{x} \rightarrow \lambda \vec{x}$, should be the way to set up the path-integral derivation of the anomaly-corrected equations. We now turn to this task. \\

\subsection{Derivation of $SO(2,1)$ anomalous equation of state}

In the previous subsection we considered scalings only in the spatial coordinates in order to make contact with Toyoda et al.'s derivation of their equation of state (what they called virial theorem), and we already saw that on account of the scale anomaly their equation had to be modified by the addition of a Jacobian piece. In order to display the anomaly of the full $SO(2,1)$ symmetry we need to consider the partition function with both time and space scaled accordingly, with the ensuing field transformation\footnote{Notice the field transformation remains the same as in the case when only the spatial components are scaled; this may have significant relevance in regularizing the Jacobian \cite{lin}.}: 

\en
Z^\lambda=\int [d\psi'^*][d\psi']e^{-S_E^{\mu' \, \beta'} [\psi'^*,\psi']},
\een

where the primed (scaled) variables are:

\begin{flalign}
&\tau \rightarrow \tau'=\lambda^2 \tau,\\ \nonumber
&\vec{x} \rightarrow \vec{x}'=\lambda \vec{x}, \\ \nonumber
&\psi(\tau,\vec{x}) \rightarrow \psi'(\tau',\vec{x}')=\lambda^{-1}\psi(\tau,\vec{x}).
\end{flalign}

Infinitesimally, for $\lambda=1+\eta$

\en
\de \tau &=2\eta \tau,\\
\de \vec{x} &=\eta \vec{x}, \\
\de \psi &=-\eta \Q(1+\vec{x}\cdot \vec{\na}+2\tau \p_\tau \W)  \psi(\tau, \vec{x}).\\
\een

Performing explicitly the scalings in the action gives (see Eq. (12) and notice that $\beta'=\lambda^2 \beta$)

\en
S_E^{\mu' , \beta'} [\psi'^*,\psi']=S_E^{\lambda^2 \mu , \beta} [\psi^*,\psi].
\een

$Z^\lambda$ can therefore be expressed as

\en
Z^\lambda &= J^\lambda \int [d\psi^*][d\psi] e^{-S_E^{\lambda^2 \mu , \beta} [\psi^*,\psi]}\\
&=J^\lambda \text{tr} \Q( e^{-\beta(H-\lambda^2 \mu N)} \W).
\een

In the large volume limit, $Z^\lambda$ becomes

\en
Z^\lambda=e^{\beta ' P' A'},
\een

where now, due to the additional rescaling of $\tau$, we expect changes, not only in the volume $A$, but also in $\beta$ and $P$. These changes are contained in both the Jacobian and the trace factors of the RHS of Eq. (47). \\

We now make an infinitesimal change $\lambda=1+\eta$ and compute the changes on the left and right hand sides of Eq(47). On the LHS:

\en
\delta Z=e^{\beta ' P' A'}-e^{\beta  P A}&=e^{\beta P A}(\beta P\delta A+PA\de \beta +\beta \pf{P}{\beta}A\de \beta) \\
&=Z(4\eta \beta PA+2\eta\beta\pf{P}{\beta}A\beta).
\een

From $PA=\f{1}{\beta}\ln Z$ we get $\beta A \pf{P}{\beta}=-PA+\f{\pf{Z}{\beta}}{Z}$, and hence,

\en
\delta Z=Z(2\eta \beta)\Q(PA+\f{\pf{Z}{\beta}}{Z} \W)=Z(2\eta \beta)(PA-\la H \ra +\mu \la N \ra),
\een

where we used $\f{1}{Z}\pf{Z}{\beta}=-\la H \ra+\mu \la N \ra$.

From the right hand side of Eq. (47) we get

\en
\delta Z=Z\Q(2\eta \beta \mu\la N \ra+2\eta \text{Tr} \Q(\hat{\theta}  \delta^3(x-y) \W)\W),
\een

where now, in view of the transformation $\delta \psi$, Eq.(45), the anomaly operator includes a time piece

\en
\hat{\theta}=-\Q(1+2\tau \p_\tau+\vec{x}\cdot \vec{\nabla} \W).
\een

Therefore, the $SO(2,1)$ anomalous equation is (as before $\ma E=\f{\la H\ra}{A}$)

\en
\ma E-P=-\f{\text{Tr} \Q(\hat{\theta} \delta^3(x-y) \W)}{\beta A}.
\een

Eqs. (26), (35) and (53) are the main results of this paper. Comparison with Eq. (8) and Eq. (3) of ref. \cite{40c} would suggest that in $2D$,  $\f{\text{Tr}\Q(\hat{\theta} \delta^3(x-y)\W)}{\beta}$ should be identified with Tan's contact term.

\section{ Comments and conclusions}

We have derived in this paper anomalous-modified equations of state for 2D scale invariant systems with contact interactions using scaling arguments and Fujikawa's path-integral approach to anomalies, Eqs. (35) and (53). In the process, as a by-product, we also obtained an anomaly-corrected virial theorem for such systems, Eq. ( 26).  The work here is of a formal nature, and it is meant to illustrate how in principle one can account for the anomalous effects in a clear fashion, which we hope will help us address some of the issues discussed in the introduction. Indeed, work based on this paper to investigate those topics is in progress, and it has already produced some concrete results (\cite{lin}; see below). A few comments are in order:

\begin{itemize}
\item While we chose the relevant 2D contact term interaction for bosonic fields as a concrete case in this paper, most of the calculations and formal manipulations rely only on the scale transformation properties of the two-body potential, $V(\lambda r)=\lambda^{-2}V(r)$. Therefore, the equations derived here also apply in the case of inverse square potentials in arbitrary spatial dimension and other $SO(2,1)$ invariant systems such as anyons \cite{56,66,67,68,69,70}. 
\item Likewise, the final results for the Eq. of state and the virial theorem apply to fermionic systems.
\item The emphasis in this paper was on the impact of anomalies in 2D, or in general, systems with a classical $SO(2,1)$ symmetry; however, most of the calculations and the ideas developed in our framework apply even when there is no such classical symmetry, or when the symmetry exists at the quantum level in some asymptotic regime (as in atoms at unitarity \cite{38}). In particular, our approach produces Eqs. similar to Eqs. (35) and (53) in $D=3$, agreeing with Toyoda et al.'s results, if one ignores the Jacobian term (see footnote 7). This procedure also seems to produce Tan's contaact term in $D=3$.
\item As it has been stated above, comparison of Eqs. (26), (35) and (53) with the literature would suggest the identification of the Jacobian contribution with Tan's contact term, and in the $2D$ case, its presence signaling the existence of an $SO(2,1)$ anomaly. In $2D$, it appears as though one need use only the spatial scaling transformation to derive the contact term, regardless of the existence of a classical $SO(2,1)$ symmetry. When dealing with the issue of a potential $SO(2,1)$ anomaly, one has to consider both time and space dilations. The operators $\hat{\theta}_s$ and $\hat{\theta}$ so obtained are different, but a proper treatment of the functional traces should produce the same answer. Further study needs to be given to this issue. 
\item It is clear that the practical relevance of the approach presented here rests upon one's ability to calculate the Jacobian, i.e, to carefully regularize it to produce usable finite (possible bare) answers. We have made progress in this direction in the bosonic sector, for the homogeneous case discussed explicitly in this paper (without the harmonic trapping term) \cite{lin}, and we have made contact with previous results in the literature that confirms the validity of our approach in this case \cite{bergman,norway}. We now give a brief summary of the salient features of the results in \cite{lin}: Following the path-integral version of Noether's theorem, and using the scaling transformations, Eq. (A.2) of the appendix, one arrives at the expression containing the anomaly term on the right hand side:

\en
\la \p_\mu j^\nu \ra=-i\,\hat{\text{T}}\text{r}\Q(\hat{\theta}\delta^3(x-y)I_2\W),
\een 

with $\hat{\theta}=\Q(-1-\vec{x}\cdot \vec{\na}-2t \p_t \W)$, $I_2=\pma 1 & 0 \\0 & 1 \epma$, and $\hat{\text{T}}\text{r}$ refers to both matrix and functional traces. \\

In the case of a constant background (consistent with the homogenous case), we can ignore the derivative contributions in $\hat{\theta}$ (notice we're using the notation $x=(x_0,\vec{x}\,)$) and regularize Eq. (54) using the eigenbasis $\phi_n$ of a Hermitian operator $M$ as follows:

\en
\delta^3(x-y) I_2=\sum \limits_{n}\phi_n(x_0,\vec{x})\phi_n^\da(y_0,\vec{y}) \rightarrow \delta^3_R(x-y) I_2=\sum \limits_{n} R\Q(\f{M}{\Lambda^2}\W)\phi_n(x_0,\vec{x})\phi_n^\da(y_0,\vec{y}),
\een

with the property that $R(0)=1$ so that at the end of the calculation we send $\Lambda \rightarrow \infty$ and $\lim \limits_{\Lambda \rightarrow \infty}R\Q(\f{M}{\Lambda^2}\W)=1$. The trace is then regularized as 

\en
\hat{\text{T}}\text{r} \Q[\delta^3_R(x-y) I_2\W]=\hat{\text{T}}\text{r} \Q[ R\Q(\f{M}{\Lambda^2}\W)\delta^3(x-y) I_2\W],
\een

with the choice of the Hermitian $M$ determined by the one-loop structure of the theory

\en
M=\begin{pmatrix}
i\p_t+\f{\nabla^2}{2}+\mu-2g\psi^* \psi+i\ep & -g \psi^2 \\
-g \psi^{*2} & -i\p_t+\f{\nabla^2}{2}+\mu-2g\psi^* \psi+i\ep
\end{pmatrix},
\een

where the fields in $M$ are constant background fields. A class of regulators of the form

\en
R\Q(\f{M}{\Lambda^2}\W)=\Q( 1 \pm \f{M}{\Lambda^2}\W)^{-1}
\een

gives, for the zero temperature case

\en
\Q\la  \p_\mu j^\mu \W\ra=-\f{g^2 (\psi^* \psi)^2}{4\pi}, 
\een

and, in the finite-temperature case considered in this paper, Eq. (53):

\en
2\ma E-2P=-\f{g^2 (\psi^* \psi)^2}{4\pi}. 
\een

In both cases we find agreement with the literature \cite{bergman,norway}. More work remains to be done, including the case when the number density is not constant, in order to make contact with other realistic situations of interest in atomic and molecular physics, as well as in condensed matter physics, and possibly in applications to holography and the AdS/CFT duality \cite{tong}. 

\end{itemize}

Acknowledgements\\

The author acknowledges productive conversations on these topics with M. Foster, K. Hazzard, and D. Sheehy. He is particularly thankful to his student, Chris Lin, not only for his help with typing the paper, but in providing constant and insightful feedback during the process of developing the framework presented here. His interactions and comments helped the author in clarifying and polishing his ideas. This paper is the final result of this enjoyable process, which will continue in the near future as we further develop the framework and apply it to physical systems of interest. This work was supported in part by the US Army Research Office Grant No. W911NF-15-1-0445.

\appendix*

\section{Jacobian contribution to Eqs. (26), (35) and (53).} 

In the change of variables, Eq. (16), we need to calculate the Jacobian

\en
J^{\lambda^{-1}}=\det 
\pma
\pfa{\psi(x)}{\psi'(y)} & \pfa{\psi(x)}{\psi'^{*}(y)}\\
\pfa{\psi^*(x)}{\psi'(y)} & \pfa{\psi^*(x)}{\psi'^{*}(y)}
\epma \equiv \det  \Q( \ma S (x,y) \W),
\een

for infinitesimal scalings $\lambda=1+\eta$. We are using the notation $x=(\tau,\vec{x}), y=(\tilde{\tau},\vec{y})$; $S(x,y)$ is the 2x2 matrix defined by Eq. (A.1). \\

The general $SO(2,1)$ infinitesimal dilation is

\en
\de \vec{x} &=\eta \vec{x}, \\
\de \tau &=2\eta \tau,\\
\de \psi &=\eta \hat{\theta}  \psi(\tau,\vec{x}),\\
\de \psi^* &=\eta \hat{\theta}  \psi^*(\tau,\vec{x}),\\
\een

where 

\en
\hat{\theta} &\equiv -\Q(1+\vec{x}\cdot \vec{\na}+2\tau \p_\tau \W).
\een

For subsection II-A, we need consider only the spatial part, $\hat{\theta}_s=-(1+\vec{x}\cdot \vec{\na})$. For these transformations $\ma S(x,y)$ becomes

\begin{flalign}
&\pma
\delta^3(x-y)-\eta \hat{\theta}_s \delta^3(x-y) & 0\\
0 & \delta^3(x-y)-\eta \hat{\theta}_s \delta^3(x-y)
\epma  \\ \nonumber
&=\pma
\delta^3(x-y)& 0\\
0 & \delta^3(x-y)
\epma +
\pma
-\eta \hat{\theta}_s \delta^3(x-y) & 0\\
0 & -\eta \hat{\theta}_s \delta^3(x-y)
\epma\\ \nonumber
&\equiv\ma I+\eta B(x,y).
\end{flalign}

In Eq.(A.4) $\de^3(x-y)=\de(\tau-\tau')\de^2(\vec{x}-\vec{y})$, $B=\pma -\hat{\theta}_s \delta^3(x-y) & 0 \\ 0 & -\hat{\theta}_s \delta^3(x-y) \epma.$ \\

We now use the well-known identity ($\hat{\text{Tr}}$ includes a functional trace as well as the 2x2 matrix trace)

\en
\det S&=e^{\hat{\text{Tr}} \ln S}\\
&=e^{\hat{\text{Tr}} \ln(\ma I+\eta  B)}\\
&=1+2 \eta \text{Tr}(B)\\
&=1-2\eta \text{Tr}\Q( \hat{\theta}_s \delta^3(x-y)\W),
\een

where $\text{Tr}$ is the functional trace only.\\

Therefore, for $\lambda=1+\eta$ ($\lambda^{-1}=1-\eta$)

\en
J^{1-\eta}=1-2\eta \text{Tr}\Q( \hat{\theta}_s \delta^3(x-y)\W).
\een

For the change of variables in the opposite direction, primed to unprimed (used in subsections II-B and II-D), we need $J^\lambda$, so the trace term in Eq. (A.6) is positive in those cases. In subsection II-D, we need to replace $\hat{\theta}_s$ by $\hat{\theta}$.


\begin{thebibliography}{46}
\expandafter\ifx\csname natexlab\endcsname\relax\def\natexlab#1{#1}\fi
\expandafter\ifx\csname bibnamefont\endcsname\relax
  \def\bibnamefont#1{#1}\fi
\expandafter\ifx\csname bibfnamefont\endcsname\relax
  \def\bibfnamefont#1{#1}\fi
\expandafter\ifx\csname citenamefont\endcsname\relax
  \def\citenamefont#1{#1}\fi
\expandafter\ifx\csname url\endcsname\relax
  \def\url#1{\texttt{#1}}\fi
\expandafter\ifx\csname urlprefix\endcsname\relax\def\urlprefix{URL }\fi
\providecommand{\bibinfo}[2]{#2}
\providecommand{\eprint}[2][]{\url{#2}}

\bibitem[{\citenamefont{Bell and Jackiw}(1969)}]{1a}
\bibinfo{author}{\bibfnamefont{J.}~\bibnamefont{Bell}} \bibnamefont{and}
  \bibinfo{author}{\bibfnamefont{R.}~\bibnamefont{Jackiw}},
  \bibinfo{journal}{Nuovo Cim.} \textbf{\bibinfo{volume}{A60}},
  \bibinfo{pages}{47} (\bibinfo{year}{1969}).

\bibitem[{\citenamefont{Adler}(1969)}]{1b}
\bibinfo{author}{\bibfnamefont{S.~L.} \bibnamefont{Adler}},
  \bibinfo{journal}{Phys.Rev.} \textbf{\bibinfo{volume}{177}},
  \bibinfo{pages}{2426} (\bibinfo{year}{1969}).

\bibitem[{\citenamefont{Adler and Bardeen}(1969)}]{1c}
\bibinfo{author}{\bibfnamefont{S.~L.} \bibnamefont{Adler}} \bibnamefont{and}
  \bibinfo{author}{\bibfnamefont{W.~A.} \bibnamefont{Bardeen}},
  \bibinfo{journal}{Phys.Rev.} \textbf{\bibinfo{volume}{182}},
  \bibinfo{pages}{1517} (\bibinfo{year}{1969}).

\bibitem[{\citenamefont{Jackiw}(1991)}]{3}
\bibinfo{author}{\bibfnamefont{R.}~\bibnamefont{Jackiw}}, \bibinfo{journal}{Beg
  Memorial Volume} \textbf{\bibinfo{volume}{A60}} (\bibinfo{year}{1991}).

\bibitem[{\citenamefont{Bonneau et~al.}(2001)\citenamefont{Bonneau, Faraut, and
  Valent}}]{a}
\bibinfo{author}{\bibfnamefont{G.}~\bibnamefont{Bonneau}},
  \bibinfo{author}{\bibfnamefont{J.}~\bibnamefont{Faraut}}, \bibnamefont{and}
  \bibinfo{author}{\bibfnamefont{G.}~\bibnamefont{Valent}},
  \bibinfo{journal}{American Journal of Physics} \textbf{\bibinfo{volume}{69}},
  \bibinfo{pages}{322} (\bibinfo{year}{2001}),
  \urlprefix\url{http://scitation.aip.org/content/aapt/journal/ajp/69/3/10.1119/1.1328351}.

\bibitem[{\citenamefont{Camblong and Ordonez}(2003)}]{4}
\bibinfo{author}{\bibfnamefont{H.~E.} \bibnamefont{Camblong}} \bibnamefont{and}
  \bibinfo{author}{\bibfnamefont{C.~R.} \bibnamefont{Ordonez}},
  \bibinfo{journal}{Phys.Rev.} \textbf{\bibinfo{volume}{D68}},
  \bibinfo{pages}{125013} (\bibinfo{year}{2003}), \eprint{hep-th/0303166}.

\bibitem[{\citenamefont{Moroz}(2011)}]{5}
\bibinfo{author}{\bibfnamefont{S.}~\bibnamefont{Moroz}},
  \bibinfo{journal}{Annals Phys.} \textbf{\bibinfo{volume}{326}},
  \bibinfo{pages}{1368} (\bibinfo{year}{2011}), \eprint{1007.4635}.

\bibitem[{\citenamefont{Camblong et~al.}(2001)\citenamefont{Camblong, Epele,
  Fanchiotti, and Garc{\'i}a~Canal}}]{b}
\bibinfo{author}{\bibfnamefont{H.~E.} \bibnamefont{Camblong}},
  \bibinfo{author}{\bibfnamefont{L.~N.} \bibnamefont{Epele}},
  \bibinfo{author}{\bibfnamefont{H.}~\bibnamefont{Fanchiotti}},
  \bibnamefont{and} \bibinfo{author}{\bibfnamefont{C.~A.}
  \bibnamefont{Garc{\'i}a~Canal}}, \bibinfo{journal}{Phys. Rev. Lett.}
  \textbf{\bibinfo{volume}{87}}, \bibinfo{pages}{220402}
  (\bibinfo{year}{2001}),
  \urlprefix\url{http://link.aps.org/doi/10.1103/PhysRevLett.87.220402}.

\bibitem[{\citenamefont{Pitaevskii and Rosch}(1997)}]{6}
\bibinfo{author}{\bibfnamefont{L.~P.} \bibnamefont{Pitaevskii}}
  \bibnamefont{and} \bibinfo{author}{\bibfnamefont{A.}~\bibnamefont{Rosch}},
  \bibinfo{journal}{Phys. Rev. A} \textbf{\bibinfo{volume}{55}},
  \bibinfo{pages}{R853} (\bibinfo{year}{1997}),
  \urlprefix\url{http://link.aps.org/doi/10.1103/PhysRevA.55.R853}.

\bibitem[{\citenamefont{Hofmann}(2012)}]{8}
\bibinfo{author}{\bibfnamefont{J.}~\bibnamefont{Hofmann}},
  \bibinfo{journal}{Phys. Rev. Lett.} \textbf{\bibinfo{volume}{108}},
  \bibinfo{pages}{185303} (\bibinfo{year}{2012}),
  \urlprefix\url{http://link.aps.org/doi/10.1103/PhysRevLett.108.185303}.

\bibitem[{\citenamefont{Tan}(2008{\natexlab{a}})}]{40a}
\bibinfo{author}{\bibfnamefont{S.}~\bibnamefont{Tan}}, \bibinfo{journal}{Annals
  of Physics} \textbf{\bibinfo{volume}{323}}, \bibinfo{pages}{2952 }
  (\bibinfo{year}{2008}{\natexlab{a}}), ISSN \bibinfo{issn}{0003-4916},
  \urlprefix\url{http://www.sciencedirect.com/science/article/pii/S0003491608000456}.

\bibitem[{\citenamefont{Tan}(2008{\natexlab{b}})}]{40b}
\bibinfo{author}{\bibfnamefont{S.}~\bibnamefont{Tan}}, \bibinfo{journal}{Annals
  of Physics} \textbf{\bibinfo{volume}{323}}, \bibinfo{pages}{2971 }
  (\bibinfo{year}{2008}{\natexlab{b}}), ISSN \bibinfo{issn}{0003-4916},
  \urlprefix\url{http://www.sciencedirect.com/science/article/pii/S0003491608000432}.

\bibitem[{\citenamefont{Tan}(2008{\natexlab{c}})}]{40c}
\bibinfo{author}{\bibfnamefont{S.}~\bibnamefont{Tan}}, \bibinfo{journal}{Annals
  of Physics} \textbf{\bibinfo{volume}{323}}, \bibinfo{pages}{2987 }
  (\bibinfo{year}{2008}{\natexlab{c}}), ISSN \bibinfo{issn}{0003-4916},
  \urlprefix\url{http://www.sciencedirect.com/science/article/pii/S0003491608000420}.

\bibitem[{\citenamefont{Olshanii et~al.}(2010)\citenamefont{Olshanii, Perrin,
  and Lorent}}]{42}
\bibinfo{author}{\bibfnamefont{M.}~\bibnamefont{Olshanii}},
  \bibinfo{author}{\bibfnamefont{H.}~\bibnamefont{Perrin}}, \bibnamefont{and}
  \bibinfo{author}{\bibfnamefont{V.}~\bibnamefont{Lorent}},
  \bibinfo{journal}{Phys. Rev. Lett.} \textbf{\bibinfo{volume}{105}},
  \bibinfo{pages}{095302} (\bibinfo{year}{2010}),
  \urlprefix\url{http://link.aps.org/doi/10.1103/PhysRevLett.105.095302}.

\bibitem[{\citenamefont{Valiente et~al.}(2011)\citenamefont{Valiente, Zinner,
  and M\o{}lmer}}]{50}
\bibinfo{author}{\bibfnamefont{M.}~\bibnamefont{Valiente}},
  \bibinfo{author}{\bibfnamefont{N.~T.} \bibnamefont{Zinner}},
  \bibnamefont{and}
  \bibinfo{author}{\bibfnamefont{K.}~\bibnamefont{M\o{}lmer}},
  \bibinfo{journal}{Phys. Rev. A} \textbf{\bibinfo{volume}{84}},
  \bibinfo{pages}{063626} (\bibinfo{year}{2011}),
  \urlprefix\url{http://link.aps.org/doi/10.1103/PhysRevA.84.063626}.

\bibitem[{\citenamefont{Fujiwara et~al.}(1993)\citenamefont{Fujiwara, Igarashi,
  Kubo, and Maeda}}]{51}
\bibinfo{author}{\bibfnamefont{T.}~\bibnamefont{Fujiwara}},
  \bibinfo{author}{\bibfnamefont{Y.}~\bibnamefont{Igarashi}},
  \bibinfo{author}{\bibfnamefont{J.}~\bibnamefont{Kubo}}, \bibnamefont{and}
  \bibinfo{author}{\bibfnamefont{K.}~\bibnamefont{Maeda}},
  \bibinfo{journal}{Nucl.Phys.} \textbf{\bibinfo{volume}{B391}},
  \bibinfo{pages}{211} (\bibinfo{year}{1993}), \eprint{hep-th/9210038}.

\bibitem[{\citenamefont{Fosco and Trinchero}(1990)}]{52}
\bibinfo{author}{\bibfnamefont{C.}~\bibnamefont{Fosco}} \bibnamefont{and}
  \bibinfo{author}{\bibfnamefont{R.}~\bibnamefont{Trinchero}},
  \bibinfo{journal}{Phys.Rev.} \textbf{\bibinfo{volume}{D41}},
  \bibinfo{pages}{1216} (\bibinfo{year}{1990}).

\bibitem[{\citenamefont{Feld and et~al}(2011)}]{53}
\bibinfo{author}{\bibfnamefont{M.}~\bibnamefont{Feld}} \bibnamefont{and}
  \bibinfo{author}{\bibnamefont{et~al}}, \bibinfo{journal}{Nature}
  \textbf{\bibinfo{volume}{480}}, \bibinfo{pages}{75} (\bibinfo{year}{2011}).

\bibitem[{\citenamefont{Petrov et~al.}(2000)\citenamefont{Petrov, Holzmann, and
  Shlyapnikov}}]{54}
\bibinfo{author}{\bibfnamefont{D.}~\bibnamefont{Petrov}},
  \bibinfo{author}{\bibfnamefont{M.}~\bibnamefont{Holzmann}}, \bibnamefont{and}
  \bibinfo{author}{\bibfnamefont{G.}~\bibnamefont{Shlyapnikov}},
  \bibinfo{journal}{Phys.Rev.Lett.} \textbf{\bibinfo{volume}{84}},
  \bibinfo{pages}{2551} (\bibinfo{year}{2000}).

\bibitem[{\citenamefont{Baur et~al.}(2012)\citenamefont{Baur, Fr\"ohlich, Feld,
  Vogt, Pertot, Koschorreck, and K\"ohl}}]{55}
\bibinfo{author}{\bibfnamefont{S.~K.} \bibnamefont{Baur}},
  \bibinfo{author}{\bibfnamefont{B.}~\bibnamefont{Fr\"ohlich}},
  \bibinfo{author}{\bibfnamefont{M.}~\bibnamefont{Feld}},
  \bibinfo{author}{\bibfnamefont{E.}~\bibnamefont{Vogt}},
  \bibinfo{author}{\bibfnamefont{D.}~\bibnamefont{Pertot}},
  \bibinfo{author}{\bibfnamefont{M.}~\bibnamefont{Koschorreck}},
  \bibnamefont{and} \bibinfo{author}{\bibfnamefont{M.}~\bibnamefont{K\"ohl}},
  \bibinfo{journal}{Phys. Rev. A} \textbf{\bibinfo{volume}{85}},
  \bibinfo{pages}{061604} (\bibinfo{year}{2012}),
  \urlprefix\url{http://link.aps.org/doi/10.1103/PhysRevA.85.061604}.

\bibitem[{\citenamefont{Levinsen and Parish}(2014)}]{58}
\bibinfo{author}{\bibfnamefont{J.}~\bibnamefont{Levinsen}} \bibnamefont{and}
  \bibinfo{author}{\bibfnamefont{M.}~\bibnamefont{Parish}}
  (\bibinfo{year}{2014}), \eprint{arXiv:1408.2737}.

\bibitem[{\citenamefont{Vogt et~al.}(2012)\citenamefont{Vogt, Feld, Frohlich,
  Pertot, Koschorreck et~al.}}]{7}
\bibinfo{author}{\bibfnamefont{E.}~\bibnamefont{Vogt}},
  \bibinfo{author}{\bibfnamefont{M.}~\bibnamefont{Feld}},
  \bibinfo{author}{\bibfnamefont{B.}~\bibnamefont{Frohlich}},
  \bibinfo{author}{\bibfnamefont{D.}~\bibnamefont{Pertot}},
  \bibinfo{author}{\bibfnamefont{M.}~\bibnamefont{Koschorreck}},
  \bibnamefont{et~al.}, \bibinfo{journal}{Phys.Rev.Lett.}
  \textbf{\bibinfo{volume}{108}}, \bibinfo{pages}{070404}
  (\bibinfo{year}{2012}), \eprint{1111.1173}.

\bibitem[{\citenamefont{Chafin and Sch\"afer}(2013)}]{62}
\bibinfo{author}{\bibfnamefont{C.}~\bibnamefont{Chafin}} \bibnamefont{and}
  \bibinfo{author}{\bibfnamefont{T.}~\bibnamefont{Sch\"afer}},
  \bibinfo{journal}{Phys. Rev. A} \textbf{\bibinfo{volume}{88}},
  \bibinfo{pages}{043636} (\bibinfo{year}{2013}),
  \urlprefix\url{http://link.aps.org/doi/10.1103/PhysRevA.88.043636}.

\bibitem[{\citenamefont{Huang}(1987)}]{65}
\bibinfo{author}{\bibfnamefont{K.}~\bibnamefont{Huang}},
  \emph{\bibinfo{title}{Statistical Mechanics}} (\bibinfo{publisher}{John Wiley
  \& Sons}, \bibinfo{year}{1987}), \bibinfo{edition}{2nd} ed.

\bibitem[{\citenamefont{Fujikawa}(1979)}]{10a}
\bibinfo{author}{\bibfnamefont{K.}~\bibnamefont{Fujikawa}},
  \bibinfo{journal}{Phys.Rev.Lett.} \textbf{\bibinfo{volume}{42}},
  \bibinfo{pages}{1195} (\bibinfo{year}{1979}).

\bibitem[{\citenamefont{Fujikawa}(1980)}]{10b}
\bibinfo{author}{\bibfnamefont{K.}~\bibnamefont{Fujikawa}},
  \bibinfo{journal}{Phys.Rev.} \textbf{\bibinfo{volume}{D21}},
  \bibinfo{pages}{2848} (\bibinfo{year}{1980}).

\bibitem[{\citenamefont{Ananos et~al.}(2003)\citenamefont{Ananos, Camblong,
  Gorrichategui, Hernadez, and Ordonez}}]{Ananos:2002id}
\bibinfo{author}{\bibfnamefont{G.~N.~J.} \bibnamefont{Ananos}},
  \bibinfo{author}{\bibfnamefont{H.~E.} \bibnamefont{Camblong}},
  \bibinfo{author}{\bibfnamefont{C.}~\bibnamefont{Gorrichategui}},
  \bibinfo{author}{\bibfnamefont{E.}~\bibnamefont{Hernadez}}, \bibnamefont{and}
  \bibinfo{author}{\bibfnamefont{C.~R.} \bibnamefont{Ordonez}},
  \bibinfo{journal}{Phys. Rev.} \textbf{\bibinfo{volume}{D67}},
  \bibinfo{pages}{045018} (\bibinfo{year}{2003}), \eprint{hep-th/0205191}.

\bibitem[{\citenamefont{Dudas and Pirjol}(1991)}]{D}
\bibinfo{author}{\bibfnamefont{E.}~\bibnamefont{Dudas}} \bibnamefont{and}
  \bibinfo{author}{\bibfnamefont{D.}~\bibnamefont{Pirjol}},
  \bibinfo{journal}{Physics Letters B} \textbf{\bibinfo{volume}{260}},
  \bibinfo{pages}{186 } (\bibinfo{year}{1991}), ISSN \bibinfo{issn}{0370-2693},
  \urlprefix\url{http://www.sciencedirect.com/science/article/pii/0370269391909894}.

\bibitem[{\citenamefont{Gaite}(2013)}]{G}
\bibinfo{author}{\bibfnamefont{J.}~\bibnamefont{Gaite}},
  \bibinfo{journal}{Phys.Usp.} \textbf{\bibinfo{volume}{56}},
  \bibinfo{pages}{919} (\bibinfo{year}{2013}), \eprint{1306.0722}.

\bibitem[{\citenamefont{Toyoda}(1993)}]{59}
\bibinfo{author}{\bibfnamefont{T.}~\bibnamefont{Toyoda}},
  \bibinfo{journal}{Phys. Rev. A} \textbf{\bibinfo{volume}{48}},
  \bibinfo{pages}{3492} (\bibinfo{year}{1993}),
  \urlprefix\url{http://link.aps.org/doi/10.1103/PhysRevA.48.3492}.

\bibitem[{\citenamefont{Toyoda and ichi Takiuchi}(1998)}]{60}
\bibinfo{author}{\bibfnamefont{T.}~\bibnamefont{Toyoda}} \bibnamefont{and}
  \bibinfo{author}{\bibfnamefont{K.}~\bibnamefont{ichi Takiuchi}},
  \bibinfo{journal}{Physica A: Statistical Mechanics and its Applications}
  \textbf{\bibinfo{volume}{261}}, \bibinfo{pages}{471 } (\bibinfo{year}{1998}),
  ISSN \bibinfo{issn}{0378-4371},
  \urlprefix\url{http://www.sciencedirect.com/science/article/pii/S0378437198003045}.

\bibitem[{\citenamefont{Takiuchi et~al.}(2000)\citenamefont{Takiuchi, Okada,
  Koizumi, Ito, and Toyoda}}]{61}
\bibinfo{author}{\bibfnamefont{K.}~\bibnamefont{Takiuchi}},
  \bibinfo{author}{\bibfnamefont{M.}~\bibnamefont{Okada}},
  \bibinfo{author}{\bibfnamefont{H.}~\bibnamefont{Koizumi}},
  \bibinfo{author}{\bibfnamefont{K.}~\bibnamefont{Ito}}, \bibnamefont{and}
  \bibinfo{author}{\bibfnamefont{T.}~\bibnamefont{Toyoda}},
  \bibinfo{journal}{Physica E: Low-dimensional Systems and Nanostructures}
  \textbf{\bibinfo{volume}{6}}, \bibinfo{pages}{810 } (\bibinfo{year}{2000}),
  ISSN \bibinfo{issn}{1386-9477},
  \urlprefix\url{http://www.sciencedirect.com/science/article/pii/S1386947799002404}.

\bibitem[{\citenamefont{Lin and Ord\'o\~nez}(2015{\natexlab{a}})}]{lin}
\bibinfo{author}{\bibfnamefont{C.~L.} \bibnamefont{Lin}} \bibnamefont{and}
  \bibinfo{author}{\bibfnamefont{C.~R.} \bibnamefont{Ord\'o\~nez}},
  \bibinfo{journal}{Phys. Rev. D} \textbf{\bibinfo{volume}{91}},
  \bibinfo{pages}{085023} (\bibinfo{year}{2015}{\natexlab{a}}),
  \urlprefix\url{http://link.aps.org/doi/10.1103/PhysRevD.91.085023}.

\bibitem[{\citenamefont{Lin and Ord\'o\~nez}(2015{\natexlab{b}})}]{virpap}
\bibinfo{author}{\bibfnamefont{C.~L.} \bibnamefont{Lin}} \bibnamefont{and}
  \bibinfo{author}{\bibfnamefont{C.~R.} \bibnamefont{Ord\'o\~nez}},
  \bibinfo{journal}{Adv. High Energy Phys.} \textbf{\bibinfo{volume}{2015}},
  \bibinfo{pages}{796275} (\bibinfo{year}{2015}{\natexlab{b}}),
  \eprint{1503.05843}.

\bibitem[{\citenamefont{Brown}(1992)}]{brown}
\bibinfo{author}{\bibfnamefont{L.}~\bibnamefont{Brown}},
  \emph{\bibinfo{title}{{Quantum field theory}}} (\bibinfo{publisher}{Cambridge
  University Press}, \bibinfo{year}{1992}).

\bibitem[{pri()}]{private}
\bibinfo{howpublished}{This was pointed out to the author by Jose Goity,
  private communication.}

\bibitem[{\citenamefont{Jackiw and Pi}(1990)}]{56}
\bibinfo{author}{\bibfnamefont{R.}~\bibnamefont{Jackiw}} \bibnamefont{and}
  \bibinfo{author}{\bibfnamefont{S.-Y.} \bibnamefont{Pi}},
  \bibinfo{journal}{Phys.Rev.} \textbf{\bibinfo{volume}{D42}},
  \bibinfo{pages}{3500} (\bibinfo{year}{1990}).

\bibitem[{\citenamefont{Comtet et~al.}(1989)\citenamefont{Comtet, Georgelin,
  and Ouvry}}]{66}
\bibinfo{author}{\bibfnamefont{A.}~\bibnamefont{Comtet}},
  \bibinfo{author}{\bibfnamefont{Y.}~\bibnamefont{Georgelin}},
  \bibnamefont{and} \bibinfo{author}{\bibfnamefont{S.}~\bibnamefont{Ouvry}},
  \bibinfo{journal}{J.Phys.} \textbf{\bibinfo{volume}{A22}},
  \bibinfo{pages}{3917} (\bibinfo{year}{1989}).

\bibitem[{\citenamefont{Arovas et~al.}(1985)\citenamefont{Arovas, Schrieffer,
  Wilczek, and Zee}}]{67}
\bibinfo{author}{\bibfnamefont{D.}~\bibnamefont{Arovas}},
  \bibinfo{author}{\bibfnamefont{J.}~\bibnamefont{Schrieffer}},
  \bibinfo{author}{\bibfnamefont{F.}~\bibnamefont{Wilczek}}, \bibnamefont{and}
  \bibinfo{author}{\bibfnamefont{A.}~\bibnamefont{Zee}},
  \bibinfo{journal}{Nucl.Phys.} \textbf{\bibinfo{volume}{B251}},
  \bibinfo{pages}{117} (\bibinfo{year}{1985}).

\bibitem[{\citenamefont{Giacconi et~al.}(1996)\citenamefont{Giacconi, Maltoni,
  and Soldati}}]{68}
\bibinfo{author}{\bibfnamefont{P.}~\bibnamefont{Giacconi}},
  \bibinfo{author}{\bibfnamefont{F.}~\bibnamefont{Maltoni}}, \bibnamefont{and}
  \bibinfo{author}{\bibfnamefont{R.}~\bibnamefont{Soldati}},
  \bibinfo{journal}{Phys.Rev.} \textbf{\bibinfo{volume}{B53}},
  \bibinfo{pages}{10065} (\bibinfo{year}{1996}).

\bibitem[{\citenamefont{Mancarella et~al.}(2013)\citenamefont{Mancarella,
  Trombettoni, and Mussardo}}]{69}
\bibinfo{author}{\bibfnamefont{F.}~\bibnamefont{Mancarella}},
  \bibinfo{author}{\bibfnamefont{A.}~\bibnamefont{Trombettoni}},
  \bibnamefont{and} \bibinfo{author}{\bibfnamefont{G.}~\bibnamefont{Mussardo}},
  \bibinfo{journal}{Nucl.Phys.} \textbf{\bibinfo{volume}{B867}},
  \bibinfo{pages}{950} (\bibinfo{year}{2013}), \eprint{1204.6656}.

\bibitem[{\citenamefont{Mancarella et~al.}(2014)\citenamefont{Mancarella,
  Mussardo, and Trombettoni}}]{70}
\bibinfo{author}{\bibfnamefont{F.}~\bibnamefont{Mancarella}},
  \bibinfo{author}{\bibfnamefont{G.}~\bibnamefont{Mussardo}}, \bibnamefont{and}
  \bibinfo{author}{\bibfnamefont{A.}~\bibnamefont{Trombettoni}},
  \bibinfo{journal}{Nucl.Phys.} \textbf{\bibinfo{volume}{B887}},
  \bibinfo{pages}{216} (\bibinfo{year}{2014}), \eprint{1407.0028}.

\bibitem[{\citenamefont{Nishida and Son}(2007)}]{38}
\bibinfo{author}{\bibfnamefont{Y.}~\bibnamefont{Nishida}} \bibnamefont{and}
  \bibinfo{author}{\bibfnamefont{D.~T.} \bibnamefont{Son}},
  \bibinfo{journal}{Phys.Rev.} \textbf{\bibinfo{volume}{D76}},
  \bibinfo{pages}{086004} (\bibinfo{year}{2007}), \eprint{0706.3746}.

\bibitem[{\citenamefont{Bergman}(1992)}]{bergman}
\bibinfo{author}{\bibfnamefont{O.}~\bibnamefont{Bergman}},
  \bibinfo{journal}{Phys.Rev.} \textbf{\bibinfo{volume}{D46}},
  \bibinfo{pages}{5474} (\bibinfo{year}{1992}).

\bibitem[{\citenamefont{Haugset and Ravndal}(1994)}]{norway}
\bibinfo{author}{\bibfnamefont{T.}~\bibnamefont{Haugset}} \bibnamefont{and}
  \bibinfo{author}{\bibfnamefont{F.}~\bibnamefont{Ravndal}},
  \bibinfo{journal}{Phys.Rev.} \textbf{\bibinfo{volume}{D49}},
  \bibinfo{pages}{4299} (\bibinfo{year}{1994}).

\bibitem[{\citenamefont{Tong}(2013)}]{tong}
\bibinfo{author}{\bibfnamefont{D.}~\bibnamefont{Tong}},
  \emph{\bibinfo{title}{Lecture notes on holographic conductivity}}
  (\bibinfo{year}{2013}).

\end{thebibliography}
\end{document}